%% file: Main.tex
\documentclass[12pt]{article}

\usepackage[linktocpage,bookmarks]{hyperref}
\usepackage[utf8]{inputenc}
\usepackage{amsmath}
\usepackage{amsthm}
\usepackage{amsfonts}          
\usepackage{amssymb}           
\usepackage{rotating}

\begin{document}

\input{Titlepage.tex}

\section{Introduction}

A recurring theme in string theory is that massless (relative to the
string scale) degrees of freedom are computed by cohomology groups of
suitable bundles or sheaves on the compactification manifold. In
general, this can be quite a difficult problem to compute and often
requires an expensive Gr\"obner basis computation when one gets down
to business. Fortunately, toric varieties are both a rather common
tool in the construction of compactification manifolds and at the same
time much more friendly for the computation of cohomology groups. The
underlying reason is the defining fact of toric varieties: there
exists an action of the algebraic torus $(\mathbb{C}^\times)^d$, where
$d$ is the dimension of the toric variety. 

For example, consider $\mathbb{P}^2$ with the line bundle
$\mathcal{O}(n)$ of first Chern class $n$. Its sections are the
homogeneous polynomials of degree $n$. However, there is a special
basis for the homogeneous polynomials, namely the homogeneous
monomials. This basis is distinguished by the weights under the torus
action. For example, let $x$, $y$, and $z$ be the homogeneous
coordinates and take the torus action to be
\begin{equation}
  (\eta, \xi) \cdot [x:y:z] = 
  \big[ x: \eta y: \xi z ]
  \qquad
  (\eta, \xi) \in (\mathbb{C}^\times)^2
\end{equation}
Then the monomial $x^a y^b z^c$ transforms with the weight $(b,c)$ and
it is the only homogeneous polynomial with this weight up to
scale. Therefore, we can identify the sections with points in the
weight lattice $\simeq \mathbb{Z}^2$. In other words, the number of
sections equals the number of integral points in the triangle $b,c\geq
0$, $b+c\leq n$. These are easy enough to count, and one finds
\begin{equation}
  \mathop{\mathrm{dim}} H^0\big(\mathbb{P}^2, \mathcal{O}(n)\big)
  =
  \frac{(n+1)(n+2)}{2} = 
  \binom{n+2}{2}
  ,\qquad
  n\geq 0
  .
\end{equation}
The same holds for sections of a line bundle $\mathcal{L}$ on a toric
variety $X_\Sigma$, each graded piece $H^0(X_\Sigma,\mathcal{L})_m$
under the torus action has multiplicity one and the allowed weights
$m\in M\simeq \mathbb{Z}^d$ form a lattice polyhedron.

For essentially the same reason, Batyrev~\cite{1993alg.geom.10003B}
was able to express the Hodge numbers of a complex 3-dimensional
Calabi-Yau hypersurface in a toric variety defined by a lattice
polytope $\nabla$ in the beautifully mirror-symmetric formula
\begin{displaymath}
  \begin{split}
    h^{11}(X_\nabla) \;&= 
    \#(\nabla)
    - 4 - 1
    - \sum_{{\mathop{\mathrm{codim}}(\nu)=1}}\mathrm{Int}(\nu) 
    + \sum_{{\mathop{\mathrm{codim}}(\nu)=2}}\mathrm{Int}(\nu)\mathrm{Int}(\nu^*)
    \\[1ex]
    h^{21}(X_\Delta) \;&= 
    \#(\Delta)
    - 4 - 1
    - \sum_{{\mathop{\mathrm{codim}}(\delta)=1}}\mathrm{Int}(\delta) 
    + \sum_{{\mathop{\mathrm{codim}}(\delta)=2}}\mathrm{Int}(\delta)\mathrm{Int}(\delta^*)
  \end{split}
\end{displaymath}
via the number of lattice points in the polytope, its dual
$\Delta=\nabla^*$, and the number of interior points in various faces.
Similar equations for complete intersections were
found~\cite{1994alg.geom.12017B, 2009arXiv0907.2701D} later as well.

\section{Counting Points}

\subsection{Naive Algorithm}

\begin{figure}
  \centerline{\includegraphics[width=0.8\textwidth]{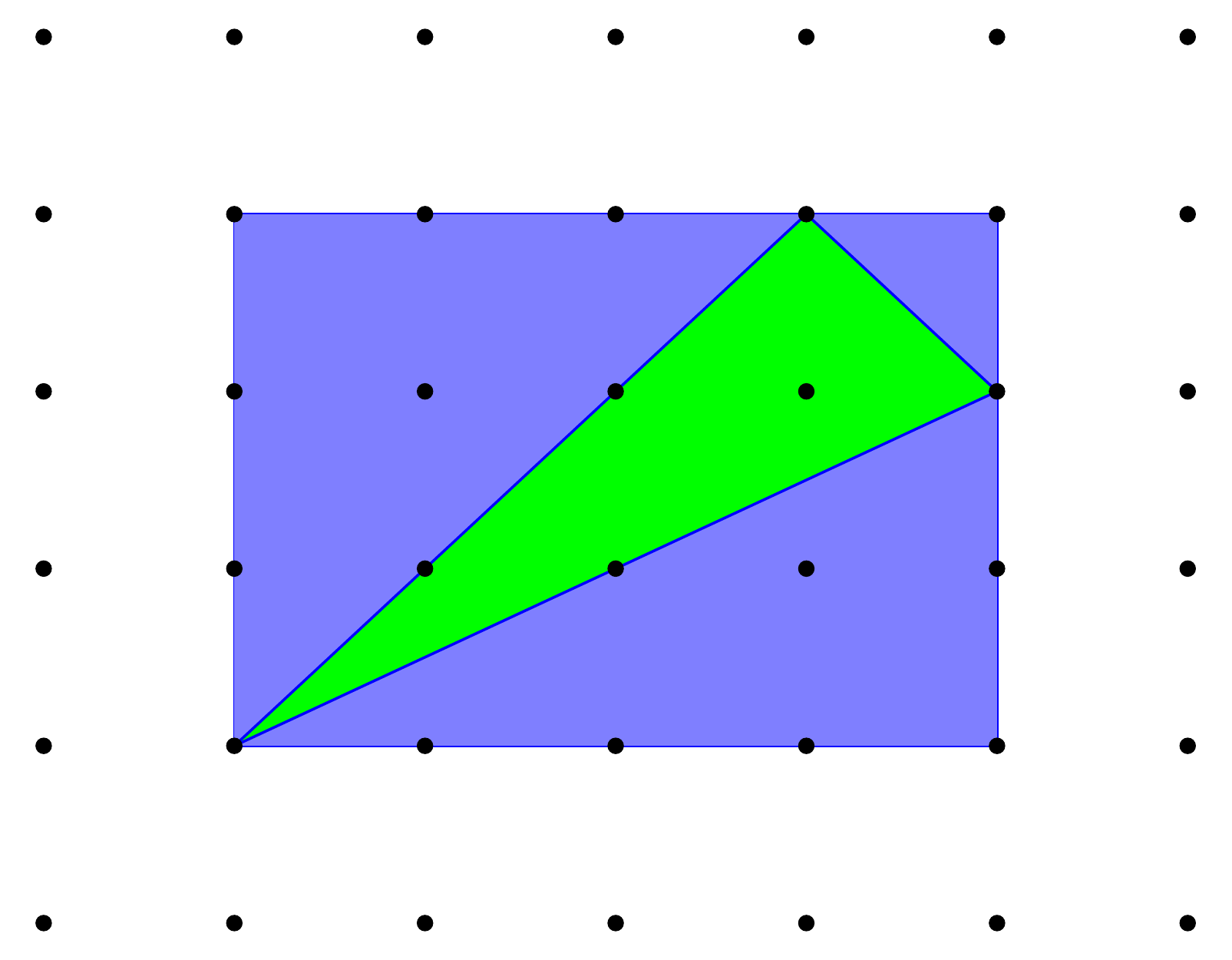}}
  \caption{Illustration of the naive point counting algorithm.} 
  \label{fig:count}
\end{figure}
For all the reasons presented in the introduction, let us now consider
the problem of enumerating the lattice points in a lattice
polytope. Note that there is also a rich story about approximate point
counts which we will completely ignore out in the following. The naive
algorithm to enumerate the points is simply to find a rectangular
bounding box (by finding the minimum and maximum values of the vertex
coordinates) and then iterate over them in a loop.  See
Figure~\ref{fig:count} for an illustration; Green is the initial
lattice polytope, blue is the bounding box. The actual loop would be
written like this in Sage:
\begin{verbatim}
sage: triangle = Polyhedron([(1,0), (0,1), (-3,-2)])
sage: pts = []
sage: for p in CartesianProduct(range(-3,1+1), range(-3,1+1)):
...       if triangle.contains(p):
...           pts.append(p)
sage: pts
[[-3, -2], [-2, -1], [-1, -1], [-1, 0], [0, 0], [0, 1], [1, 0]]
\end{verbatim}
Sage also contains PALP~\cite{2004CoPhC.157...87K,
  2011arXiv1106.4529B, 2012arXiv1205.4147B,
  Novoseltsev:lattice_polytope} and has a friendly interface for
it. The wrapper class returns the points as the columns of a matrix:
\begin{verbatim}
sage: triangle = LatticePolytope([(1,0), (0,1), (-3,-2)])
sage: triangle.points()
[ 1  0 -3 -2 -1 -1  0]
[ 0  1 -2 -1  0 -1  0]
\end{verbatim}

\subsection{Smarter Ways}

Much more can be done for counting lattice points. For one, there is a
curse of dimensionality: the volume of the $d$-dimensional unit simplex is
$\frac{1}{d!}$ times the volume of the unit hypercube. Hence, going
through all points in a bounding box becomes less and less
efficient. Note that one can directly enumerate the points of a
simplex using the Smith normal norm. This suggests to
\begin{enumerate}
\item Triangulate the lattice polytope
\item Enumerate the points in each simplex
\end{enumerate}

A yet more sophisticated way to enumerate lattice points is Barvinok's
algorithm~\cite{MR1304623} and uses generating functions. The data of
the integral points in a lattice polytope $P$ can clearly be encoded
in the polynomial
\begin{equation}
  f_P(x) = \sum_{(n_1,\dots,n_d)\in P} x_1^{n_1} \cdots x_d^{n_d}
  .
\end{equation}
In particular, $f_p(1)$ equals the number of integral points in
$P$. The central result is that this generating function can be
written as a rational function. For example, take the 1-dimensional
lattice interval $[n_1, n_2]$ with $n_1$, $n_2\in\mathbb{Z}$. Then
\begin{equation}
  f_{[n_1,n_2]}(x) 
  = \sum_{i = n_1}^{n_2} x^i
  = \frac{x^{n_1}}{1-x} - \frac{x^{n_2+1}}{1-x}
  .
\end{equation}
For any lattice polytope there exists an analogous formula by clever
combinations of arithmetic series and subtracting the overcounted
points.

\subsection{Implementation}

For purposes of toric geometry we are mainly interested in reflexive
polytopes in dimensions $3$, $4$, and perhaps $5$. These have not too
many integral points, for example a $4$-dimensional reflexive polytope
has between $6$ and $680$ lattice points~\cite{Kreuzer:2000xy} and the
average is closer to the lower bound. In practical terms, this means
that the necessary pre-processing (triangulation, Smith forms) renders
all improved algorithms actually slower than the naive
approach. However, one still needs to make some crucial optimizations
in the implementation of the naive algorithm. These are:
\begin{itemize}
\item Unwind the inner (and next-to-inner) loop, that is, rewrite 
  inequalities
  \begin{equation}
    Ax \leq b
    \quad \Leftrightarrow \quad
    a_1 x_1 ~\leq~ b - \sum_{i=2}^d a_i x_i
  \end{equation}
  such that one needs only one multiplication when iterating over
  $x_1$.
\item Permute coordinates such that the longest edge of the bounding
  box is the innermost loop.
\item Reorder inequalities to always try the most restrictive
  inequality first.
\item Use convexity: If two points are in the polytope, so are all
  intermediate points.
\end{itemize}
There is no doubt that these optimizations are implemented in PALP,
though there is no clear documentation and the source code is quite
hard to read. Furthermore, PALP is really only applicable towards
reflexive polytopes and sometimes one needs to compute with
non-reflexive polytopes for which certain compile-time assumptions in
PALP will fail. For these reasons, the author has re-implemented this
algorithm in Sage~\cite{Sage, BraunHampton:polyhedra}, so yet another
way to compute the integral points in Figure~\ref{fig:count} is
\begin{verbatim}
sage: triangle = Polyhedron([(1,0), (0,1), (-3,-2)])
sage: triangle.integral_points()
((-3, -2), (-2, -1), (-1, -1), (-1, 0), (0, 0), (1, 0), (0, 1))
\end{verbatim}
The implementation is written in Cython~\cite{Cython}, a variant of
the Python language that can be compiled to native machine code. It
automatically checks bounds and uses machine integers if possible
possible and arbitrary-precision integers when necessary. Finally,
there are no dimension or size restrictions, nor does the polytope
have to be full-dimensional or contain the origin.

\begin{figure}
  \centerline{\includegraphics[width=0.8\textwidth]{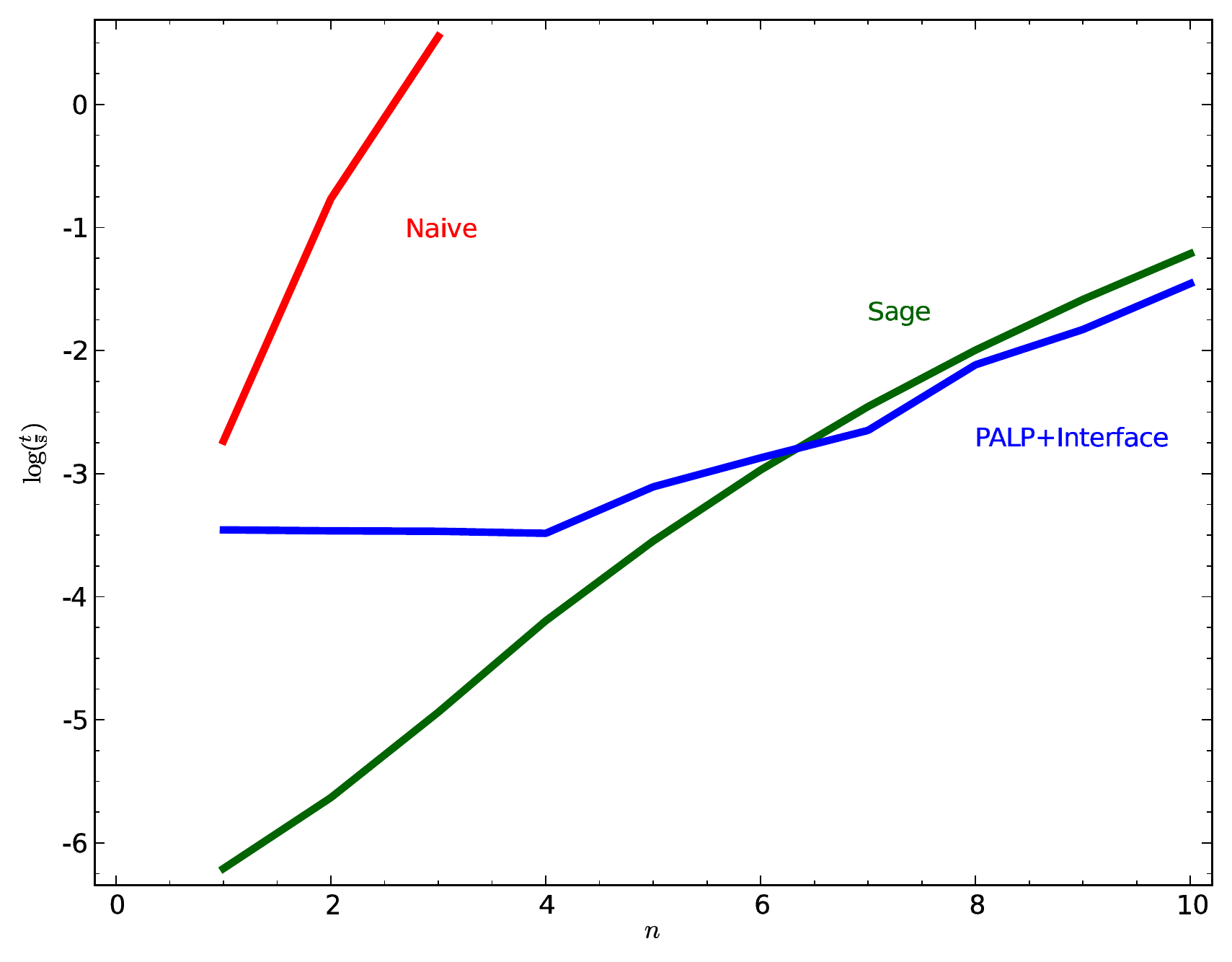}}
  \caption{Comparison of a naive implementation, PALP via a pseudo-tty
    interface, and the author's implementation in Sage.}
  \label{fig:compare}
\end{figure}

In Figure~\ref{fig:compare}, the three different implementations are
compared for the 4-dimensional cross-polytope scaled by a factor of
$n$.
Of course just running PALP without parsing the output into Sage would
be the fastest, but then that is not really useful for general
computations. As one can see from Figure~\ref{fig:compare}, for
polytopes with relatively few points (for example, reflexive ones)
trying to interface with an external program is limited by the latency
of executing the program and setting up redirections for
stdin/out. This illustrates one of the key findings of the Sage
project: in order to leverage domain-specific solutions into a
framework that combines various mathematical disciplines, it is
crucial to develop libraries and not stand-alone monolithic
programs. For example, much effort has been spent to separate
Singular~\cite{GPS05} into a shared library, and this made Singular
much more useful to a general audience outside of computational
algebraic geometry.

\section{Hilbert Polynomials}

As an application that was suggested during the Max's memorial
conference, let us compute the Erhard polynomials of the $473,800,776$
reflexive 4-dimensional polytopes and, from that, the Hilbert
polynomials of the corresponding Calabi-Yau hypersurfaces. This
illustrates the aforementioned need to combine libraries for
domain-specific problems to perform an interdisciplinary computation.
In particular, one needs to 
\begin{itemize}
\item Access the database of 4-d reflexive polytopes via PALP,
\item Compute the dual description for dilated (non-reflexive)
  polytopes, for which we use the Parma Polyhedra
  Library~\cite{BagnaraHZ08SCP},
\item Count the integral points as implemented in Sage, and
\item Polynomial algebra via {\tt libSingular}~\cite{GPS05, 
  libsingular}.
\end{itemize}

The basic question is to count the number of integral points of a
lattice polytope after dilating it by a factor of
$n\in\mathbb{Z}_{\geq}$. For example, take the polytope defining
$\mathbb{P}^4$,
\begin{equation}
  \nabla = \mathop{\mathrm{conv}}\{
  { \scriptstyle
    (1,0,0,0),~
    (0,1,0,0),~
    (0,0,1,0),~
    (0,0,0,1),~
    (-1,-1,-1,-1)
  }
  \}.
\end{equation}
By direct computation one can easily find the number of integral
points for low-lying values of $n$, namely $E(\nabla,n) = 1$, $6$,
$21$, $56$, $126$, $251$, $456$, $771$, $1231$, $1876$, $\dots$. On
general grounds, these point counts must be the values of a
polynomial, the so-called Erhard polynomial $E(\nabla,n)$. Since the
point count approximates the volume for large $n$, the Erhard
polynomial must be a polynomial of the same degree as the ambient
space. Hence, to compute the Erhard polynomial one can just pick $\geq
5$ values and compute their Lagrange polynomial. For example,
\begin{verbatim}
sage: R.<x> = QQ[]
sage: R.lagrange_polynomial([(0,1),(1,6),(2,21),(3,56),(4,126),
...       (5,251),(6,456),(7,771),(8,1231),(9,1876)])
5/24*x^4 + 5/12*x^3 + 55/24*x^2 + 25/12*x + 1
\end{verbatim}
The Erhard polynomial can also be evaluated at negative values, even
though somewhat unnatural at first sight. It turns out to be much more
symmetric if we allow negative values, for example
\begin{displaymath}
\begin{tabular}{c|ccccccccccc}
  $n$ & $-5$ & $-4$ & $-3$ & $-2$ & $-1$ & $0$ & $1$ & $2$ & $3$ & $4$ & $5$
  \\ 
  $E(\nabla,n)$ & $126$ & $56$ & $21$ & $6$ & $1$ & $1$ & $6$ & $21$ & $56$ & $126$ & $251$
\end{tabular}
\end{displaymath}
This symmetry is called Erhard reciprocity,\cite{MR2810322}
\begin{equation}
  E(P,-n) = 
  (-1)^d \#\big\{\mathop{\mathrm{Int}}(nP) \cap \mathbb{Z}^d\big\}
  = (-1)^d E(P,n-1)
  .
\end{equation}
Note that the second equality is a consequence of reflexivity of the
polytope $P$, while the first equality holds for arbitrary lattice
polytopes. Hence, for purposes of computing the Erhard polynomial of a
4-d reflexive polytope, we only need to compute two numbers: $E(P,1)$
and $E(P,2)$. The rest follows from Erhard reciprocity and the fact
that $E(P,0) = E(P,-1) = 1$. This is a quite tractable enumeration
problem, even if repeated $473,800,776$ times.

For a toric variety $X_\nabla$, the integral points of the dual
polytope $\Delta = \nabla^*$ can be identified with the monomial basis
for the sections of the anticanonical bundle, as alluded to in the
introduction. Therefore, the Hilbert polynomial of the anticanonical
bundle
\begin{equation}
  \chi(X_\nabla, -K,n) 
  =
  \mathop{\mathrm{dim}}
  H^0\big(X_\nabla, -K^{\otimes n}\big)
  =
  \#\{n \Delta \cap \mathbb{Z}^d\}
  =
  E(\Delta,n)
\end{equation}
equals the Erhard polynomial of the dual lattice polytope. Up to a
factor of $\tfrac{1}{d!}$, the leading coefficient of the Hilbert
polynomial $\chi(X_\nabla,-K,n)=\sum_0^d a_k n^k$ hence measures the
number of points in the dual polytope which equals the degree of the
variety,
\begin{equation}
  a_d = \tfrac{1}{d!} \mathop{\mathrm{deg}}(X_\nabla) = 
  \tfrac{1}{d!} \int c_1(X_\nabla)^d
\end{equation}
Since the Erhard polynomial for a 4-d reflexive polytope has two
essential degrees of freedom, this raises the question of what the
other geometric quantity is encoding. This turns out to have a nice
answer, it is the average scalar curvature of the variety.

Finally, physicists are of course mostly interested in the Calabi-Yau
hypersurfaces $Y_\nabla \subset X_\nabla$ inside toric varieties. By a
standard computation this must be an anticanonical hypersurface. Using
the long exact sequence for the restriction to the hypersurface, one
obtains the Hilbert polynomial
\begin{equation}
  \label{eq:HilbertSeries}
  \begin{split}
    \chi(Y_\nabla,n) \;&= 
    H^0(X_\nabla,K^n) - H^0(X_\nabla,K^{n-1})
    \\ &=
    \frac{\mathop{\mathrm{deg}}(Y_\nabla)}{d!} 
    n^d
    + 0 + a_{d-2} n^{d-2} + \cdots + a_1 n + 0
  \end{split}
\end{equation}
where we used the vanishing of the average scalar curvature and of the
arithmetic genus to eliminate $a_{d-1}$ and $a_0$. Hence, the Hilbert
polynomial 
\begin{equation}
  \chi(Y_\nabla, \pi^*K,n) =  H(X_\nabla,n) - H(X_\nabla,n-1)
  = a_3 n^3 + a_1 n
\end{equation}
of the polarized Calabi-Yau threefold $(Y_\nabla,\pi^* K)$ has only
two non-vanishing coefficients.

\section{Results}

\begin{figure}[tb]
  \centerline{\includegraphics[width=\textwidth]{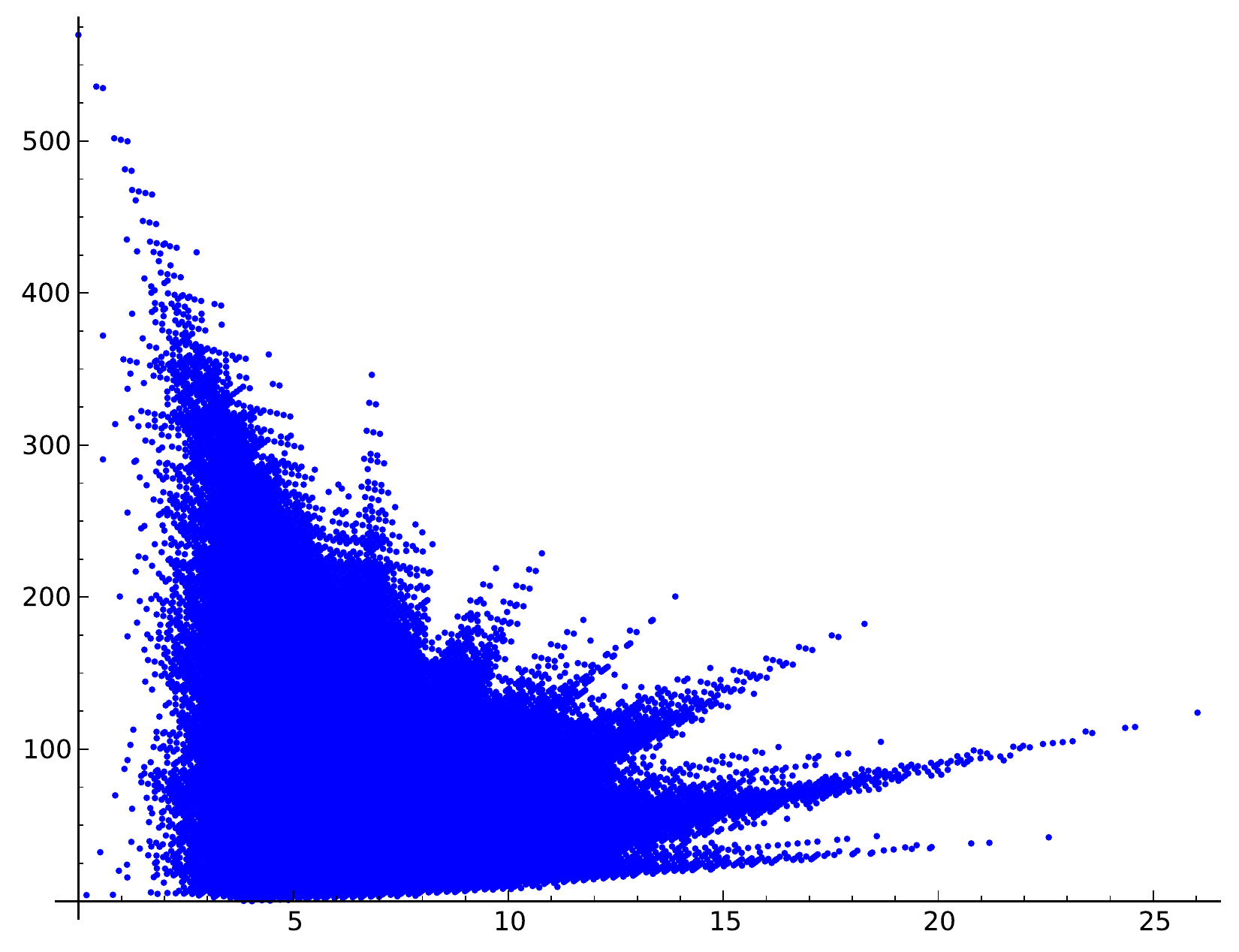}}
  \caption{The points $(a_1 - \tfrac{91}{588}a_3,
    a_3-\tfrac{1}{5}a_1)$ corresponding to the $14,373$ distinct
    Hilbert series data $(a_1,a_3)$. The linear transformation is
    chosen in order to fill the positive quadrant.}
  \label{fig:Hilbert}
\end{figure}
\begin{sidewaysfigure}[p]
  \centerline{
    \includegraphics[width=0.45\textwidth]{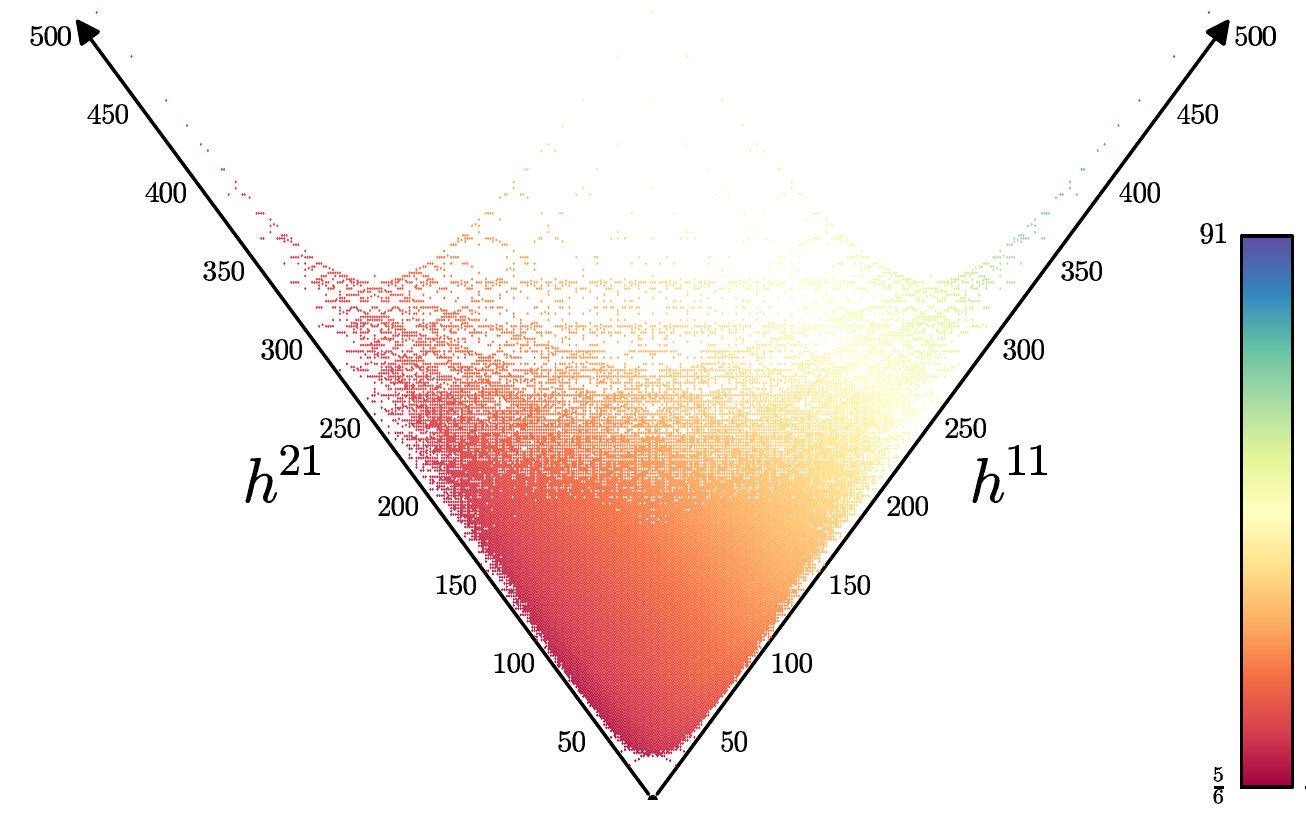}
    \includegraphics[width=0.45\textwidth]{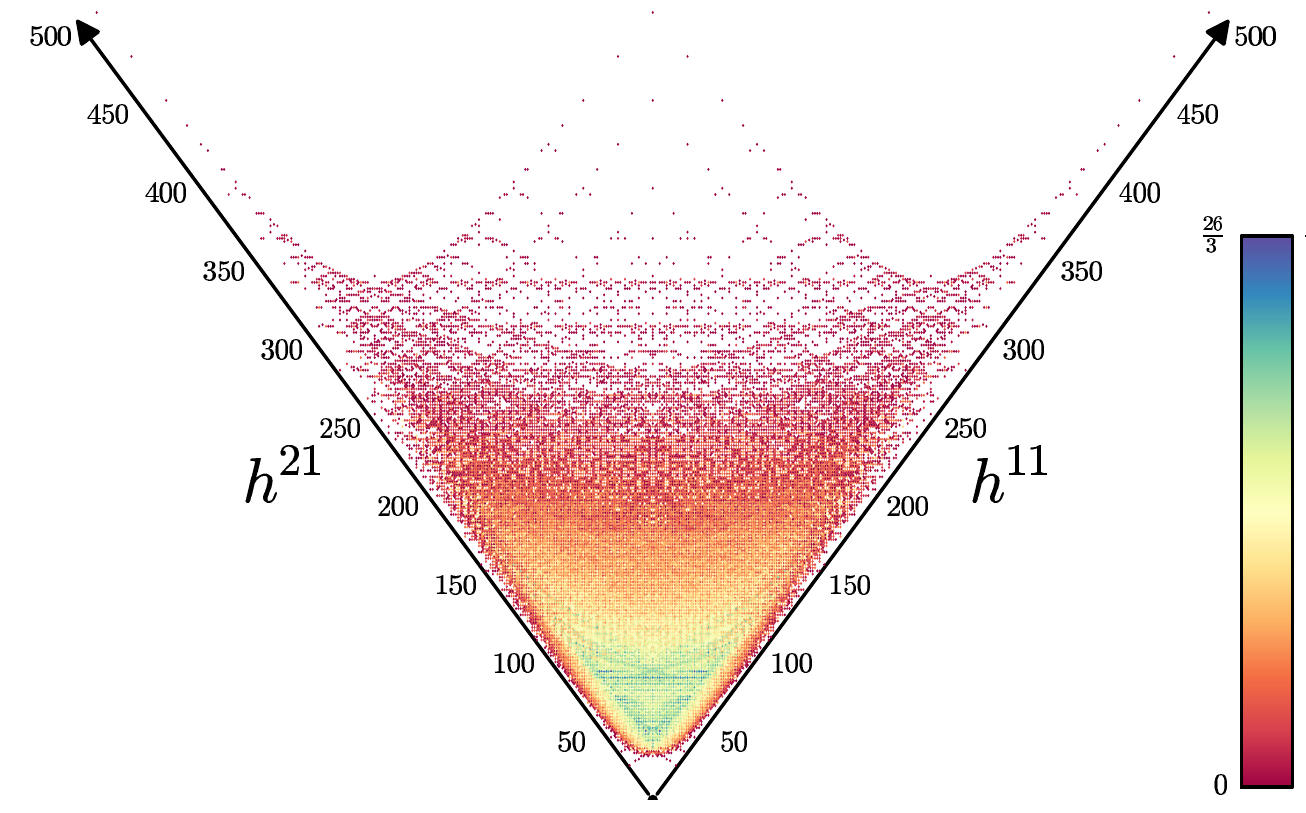}
  }
  \vspace{4mm}
  \centerline{
    \includegraphics[width=0.45\textwidth]{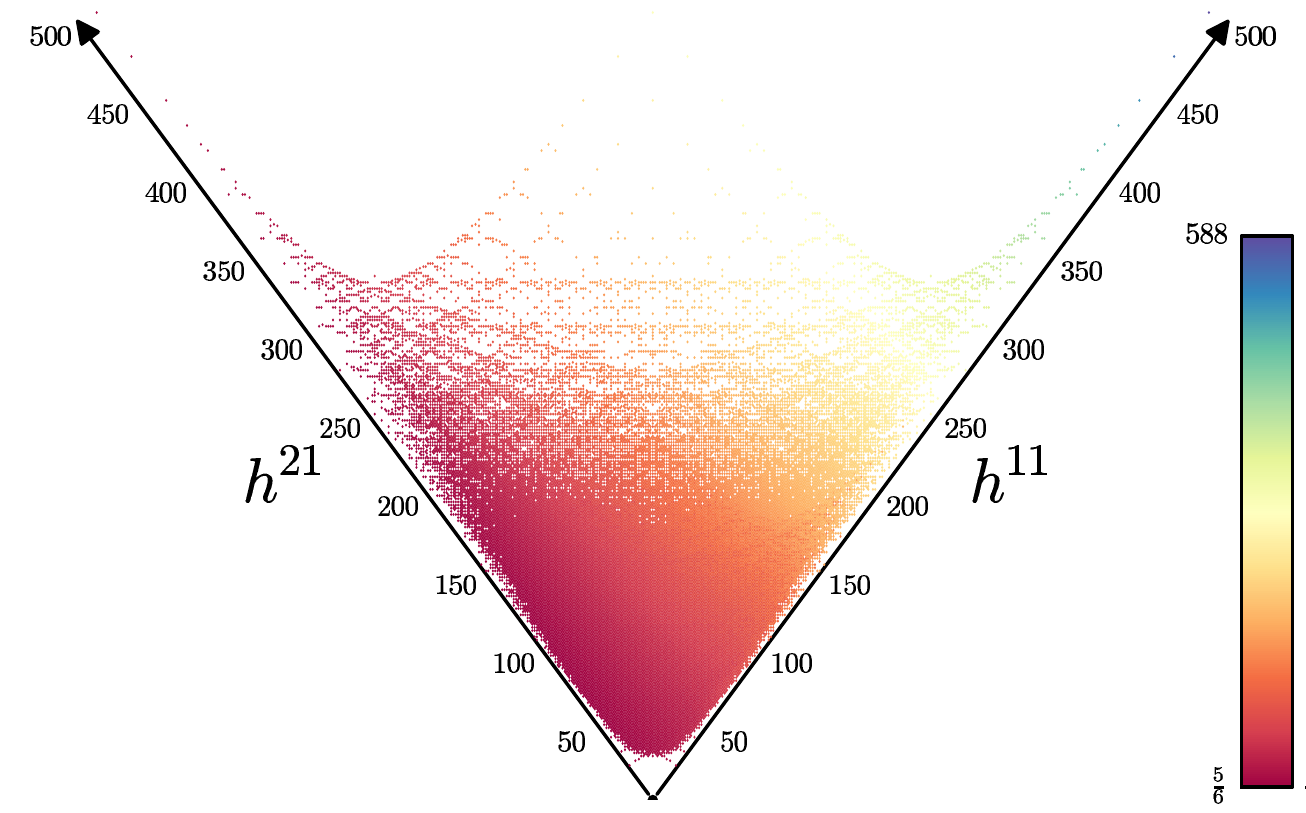}
    \includegraphics[width=0.45\textwidth]{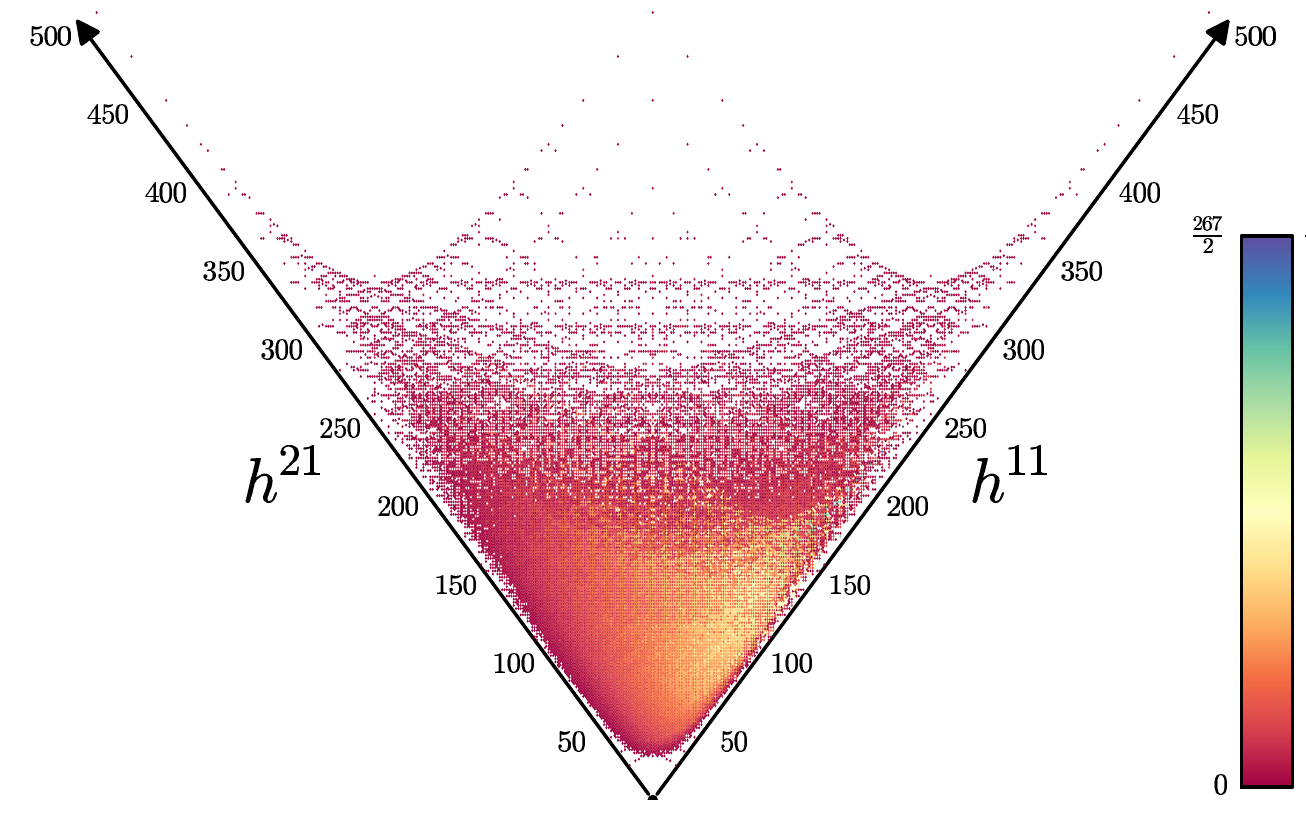}
  }
  \caption{Overview over the Hilbert series coefficients $(a_1,a_3)$
    for each pair of Hodge numbers $(h^{11},h^{21})$. In the left
    column, the average $(a_i^\text{max}+a_i^\text{min})/2$ between
    the largest and the smallest coefficient. In the right column, the
    difference $a_i^\text{max}-a_i^\text{min}$. The top row shows
    $a_1$, the bottom row shows $a_3$.  }
  \label{fig:Plot}
\end{sidewaysfigure}
According to eq.~\eqref{eq:HilbertSeries}, the Hilbert series of a
polarized Calabi-Yau threefold $(Y_\nabla,\pi^* K)$ boils down to a
pair of numbers $(a_3,a_1) \in \mathbb{Q}^2$. In fact, there are
$14,373$ distinct Hilbert series with $386$ different values for
$\tfrac{5}{6}\leq a_1 \leq 91$ and $2,229$ different values for
$\tfrac{5}{6}\leq a_3 \leq 588$. Moreover, the Hilbert series
coefficients lie in a fairly narrow wedge $\tfrac{1}{5} \leq
\tfrac{a_3}{a_1} \leq \tfrac{84}{13} = \tfrac{588}{91}$ of the
$\mathbb{Q}^2$ plane. In Figure~\ref{fig:Hilbert}, we stretch this
wedge to fill the positive quadrant.

It probably comes as no surprise that there is a correlation between
the Hilbert series data and the $30,108$ Hodge pairs $(h^{11},
h^{21})$. In fact, the Hilbert polynomial coefficient are roughly
proportional to $h^{11}$. For example, the maximal value
$(a_1,a_3)=(91,588)$ is attained at the manifold with
$(h^{11},h^{21})=(491,11)$ with the largest known $h^{11}$. This
dependence on $h^{11}$ is nicely illustrated by the left column in
Figure~\ref{fig:Plot}. However, while this overall tendency is clearly
visible, note that there is no precise relation. For most Hodge pairs,
there are multiple allowed values for the Hilbert series
coefficients. As can be seen in the right column of
Figure~\ref{fig:Plot}, the spread in $a_1$ is qualitatively different
from the spread in the $a_3$ coefficient. At this point, the author
has no mathematical explanation for this behavior. The Hilbert series
data is available online~\cite{datafile}.

\section{Conclusions}

The Kreuzer-Skarke enumeration of reflexive 4-dimensional polytopes
remains the largest single effort in what could be called
computational string theory. When it was performed, it was an amazing
feat relative to the available processing power. For example, the
authors were not able to hold all enumerated polytopes in memory and
used a carefully bit-packed hard drive cache. But technology improved
by leaps and bounds in the meantime; In 2011, the requisite amount of
RAM costs $20\$$. Today, we can finally \emph{do something} with this
giant database. Moreover, we no longer need to write hand-crafted C
code but can use a mix of interpreted languages and existing
libraries, increasing both maintainability and code reuse. And one of
these building blocks is and remains PALP, which was written before
most of the tools we can rely on today were created.

\bibliographystyle{utcaps}
\bibliography{VolkerBraun}

\end{document}

%% file: Titlepage.tex
\begin{titlepage}
  \vspace*{-2cm}
  \hfill
  \parbox[c]{5cm}{
    \begin{flushright}
    \end{flushright}
  }
  \vspace*{2cm}
  \begin{center}
    \Huge 
    Counting Points and Hilbert Series\\
    in String Theory
  \end{center}
  \vspace*{8mm}
  \begin{center}
    \begin{minipage}{\textwidth}
      \begin{center}
        \sc 
        Volker Braun
      \end{center}
      \begin{center}
        \textit{
          Dublin Institute for Advanced Studies\hphantom{${}^1$}\\
          10 Burlington Road\\
          Dublin 4, Ireland
        }
      \end{center}
      \begin{center}
        \texttt{Email: vbraun@stp.dias.ie}
      \end{center}
    \end{minipage}
  \end{center}
  \vspace*{\stretch1}
  \begin{abstract}
  The problem of counting points is revisited from the perspective of
  reflexive 4-dimensional polytopes. As an application, the Hilbert
  series of the $473,800,776$ reflexive polytopes (equivalently, their
  Calabi-Yau hypersurfaces) are computed.
  \end{abstract}
  \vspace*{\stretch1}
\end{titlepage}
\tableofcontents
\listoffigures 	


%% file: Main.bbl
\providecommand{\href}[2]{#2}\begingroup\raggedright\begin{thebibliography}{10}

\bibitem{1993alg.geom.10003B}
V.~V. {Batyrev}, ``{Dual Polyhedra and Mirror Symmetry for Calabi-Yau
  Hypersurfaces in Toric Varieties},'' in {\em eprint arXiv:alg-geom/9310003},
  p.~10003.
\newblock Oct., 1993.

\bibitem{1994alg.geom.12017B}
V.~V. {Batyrev} and L.~A. {Borisov}, ``{On Calabi-Yau Complete Intersections in
  Toric Varieties},'' in {\em eprint arXiv:alg-geom/9412017}, p.~12017.
\newblock Dec., 1994.

\bibitem{2009arXiv0907.2701D}
C.~F. {Doran} and A.~Y. {Novoseltsev}, ``{Closed form expressions for Hodge
  numbers of complete intersection Calabi-Yau threefolds in toric varieties},''
  {\em ArXiv e-prints} (July, 2009) \href{http://arXiv.org/abs/0907.2701}{{\tt
  0907.2701}}.

\bibitem{2004CoPhC.157...87K}
M.~{Kreuzer} and H.~{Skarke}, ``{PALP: A Package for Analysing Lattice
  Polytopes with applications to toric geometry},'' {\em Computer Physics
  Communications} {\bf 157} (Feb., 2004) 87--106,
  \href{http://arXiv.org/abs/arXiv:math/0204356}{{\tt arXiv:math/0204356}}.

\bibitem{2011arXiv1106.4529B}
A.~P. {Braun} and N.-O. {Walliser}, ``{A new offspring of PALP},'' {\em ArXiv
  e-prints} (June, 2011) \href{http://arXiv.org/abs/1106.4529}{{\tt
  1106.4529}}.

\bibitem{2012arXiv1205.4147B}
A.~P. {Braun}, J.~{Knapp}, E.~{Scheidegger}, H.~{Skarke}, and N.-O. {Walliser},
  ``{PALP - a User Manual},'' {\em ArXiv e-prints} (May, 2012)
  \href{http://arXiv.org/abs/1205.4147}{{\tt 1205.4147}}.

\bibitem{Novoseltsev:lattice_polytope}
A.~Y. Novoseltsev, {\em The {\tt lattice\_polytope} module of {S}age}.
\newblock The Sage Development Team, 2011.
\newblock
  \url{http://www.sagemath.org/doc/reference/sage/geometry/lattice\_polytope}.

\bibitem{MR1304623}
A.~I. Barvinok, ``A polynomial time algorithm for counting integral points in
  polyhedra when the dimension is fixed,'' {\em Math. Oper. Res.} {\bf 19}
  (1994), no.~4, 769--779.

\bibitem{Kreuzer:2000xy}
M.~Kreuzer and H.~Skarke, ``{Complete classification of reflexive polyhedra in
  four-dimensions},'' {\em Adv.Theor.Math.Phys.} {\bf 4} (2002) 1209--1230,
  \href{http://arXiv.org/abs/hep-th/0002240}{{\tt hep-th/0002240}}.

\bibitem{Sage}
W.~A. Stein {\em et al.}, {\em {S}age {M}athematics {S}oftware ({V}ersion
  4.7)}.
\newblock The Sage Development Team, 2011.
\newblock \url{http://www.sagemath.org}.

\bibitem{BraunHampton:polyhedra}
V.~Braun and M.~Hampton, {\em The {\tt polyhedra} module of {S}age}.
\newblock The Sage Development Team, 2011.
\newblock \url{http://sagemath.org/doc/reference/sage/geometry/polyhedra.html}.

\bibitem{Cython}
G.~Ewing, R.~W. Bradshaw, S.~Behnel, D.~S. Seljebotn, {\em et al.}, {\em
  {C}ython compiler ({V}ersion 0.14)}, 2011.
\newblock \url{http://www.cython.org}.

\bibitem{GPS05}
G.-M. Greuel, G.~Pfister, and H.~Sch\"onemann, ``{\sc Singular} 3.1.3,'' a
  computer algebra system for polynomial computations, Centre for Computer
  Algebra, University of Kaiserslautern, 2011.
\newblock \url{http://www.singular.uni-kl.de}.

\bibitem{BagnaraHZ08SCP}
R.~Bagnara, P.~M. Hill, and E.~Zaffanella, ``The {Parma Polyhedra Library}:
  Toward a Complete Set of Numerical Abstractions for the Analysis and
  Verification of Hardware and Software Systems,'' {\em Science of Computer
  Programming} {\bf 72} (2008), no.~1--2, 3--21.

\bibitem{libsingular}
M.~Albrecht, ``{\tt libSingular}.''
  \url{https://groups.google.com/forum/#!forum/libsingular-devel}.

\bibitem{MR2810322}
D.~A. Cox, J.~B. Little, and H.~K. Schenck, {\em Toric varieties}, vol.~124 of
  {\em Graduate Studies in Mathematics}.
\newblock American Mathematical Society, Providence, RI, 2011.

\bibitem{datafile}
V.~Braun,
  ``\url{http://www.stp.dias.ie/\textasciitilde{}vbraun/reflexive4d/Hilbert\textunderscore{}Hodge.sobj},''
  2011.

\end{thebibliography}\endgroup
